\documentclass[12pt]{article}
\usepackage{epsf,graphics,epsfig,graphicx}

\newcommand{\lsim}{\stackrel{<}{_\sim}}
\newcommand{\gsim}{\stackrel{>}{_\sim}}

\def\be{\begin{equation}}
\def\ee{\end{equation}}
\def\bea{\begin{eqnarray}}
\def\eea{\end{eqnarray}}

\begin{document}

\begin{titlepage}

\begin{flushright}
LMU 04/04\\
hep-ph/0404220\\
May 2004\\
\end{flushright}

\vspace{0.5cm}
\begin{center}
\Large\bf\boldmath
$b$-physics signals of the lightest CP-odd Higgs in the NMSSM at
large $\tan \beta$\\
\unboldmath
\end{center}

\vspace{0.8cm}
\begin{center}

Gudrun Hiller$^{1}$\\[0.1cm]
{\sl
Ludwig-Maximilians-Universit\"at M\"unchen, Sektion Physik, 
Theresienstra\ss{}e 37 \\ D-80333 M\"unchen, Germany}\\[0.4cm]

\end{center}

\footnotetext[1]{email: hiller@theorie.physik.uni-muenchen.de}

\vspace{0.3cm}
\begin{abstract}
\vspace{0.1cm}
\noindent

We investigate the low energy phenomenology of the lighter pseudoscalar 
$A_1^0$ in the NMSSM. The $A_1^0$ mass can naturally 
be small due to a global $U(1)_R$ symmetry of the Higgs potential, which 
is only broken by trilinear soft terms.
The $A_1^0$ mass  is further 
protected from renormalization group effects in the large $\tan \beta$ limit.
We calculate the $b \to s A_1^0$ amplitude at leading order in $\tan \beta$
and work out the contributions to rare $K$, $B$ and 
radiative $\Upsilon$-decays and $B -\bar B$ mixing.
We obtain constraints on the $A_1^0$ mass and couplings 
and show that masses down to ${\cal{O}}(10)$ MeV are allowed.
The $b$-physics phenomenology  of the NMSSM differs from the MSSM in 
the appearance of sizeable renormalization effects 
from neutral Higgses to the photon and gluon dipole operators
and the breakdown of the MSSM 
correlation between the $B_s \to \mu^+ \mu^-$ branching ratio
and $B_s - \bar B_s$ mixing.
For  $A_1^0$ masses above the tau threshold the $A_1^0$ can be searched for
in $b \to s \tau^+ \tau^-$ processes with
branching ratios $\lsim 10^{-3}$.

\end{abstract}

\vspace{0.4 cm}

\end{titlepage}

\section{Introduction}

Sizeable flavor changing neutral current (FCNC) effects in meson decays 
arise in the minimal supersymmetric Standard Model 
(MSSM) at large $\tan \beta$, e.g.~\cite{Babu:1999hn}-\cite{Buras:2002vd}.
In this model, the amplitude of exchanging neutral Higgses
between down-type fermions $f$, i.e.~down-type quarks or charged leptons
\be
\sum_{S=h^0,H^0,A^0} \frac{(g_{\bar f f S})^2}{m_S^2} \propto
-\frac{\cos^2 (\beta-\alpha)}{m_{h^0}^2} -
\frac{\sin^2 ( \beta-\alpha)}{m_{H^0}^2}+\frac{1}{m_{A^0}^2} = 0
\label{eq:tree}
\ee
vanishes. 
Here, $m_S$, $g_{\bar f f S}$ denote the Higgs masses and couplings to
a fermion pair, respectively and $\alpha$ is the
scalar mixing angle. Eq.~(\ref{eq:tree}) implies that the Wilson coefficients
for $b \to s \ell^+ \ell^-$ decays from scalar and pseudoscalar boson
exchange in the MSSM at large $\tan \beta$ 
are equal with opposite sign \cite{Huang:2000sm},\cite{Bobeth:2001sq}.
If the relation is broken, interesting effects via operator mixing are
induced \cite{Hiller:2003js}. In particular, the dipole operators
responsible for $b \to  s \gamma$ and $b \to s g$ decays receive
sizeable contributions from the neutral Higgs bosons.
Furthermore, specific contributions to $B -\bar B$ mixing from
scalar exchange arise.
This happens in the presence of more Higgses,
such as in the next-to-minimal supersymmetric Standard Model (NMSSM).

The NMSSM is the MSSM extended by a singlet $N$, with
the superpotential \cite{Nilles:1982dy,Ellis:1988er}
\be
W=Q Y_u H_u U+ Q Y_d H_d D+ L Y_e H_d E +
\lambda H_d H_u N -\frac{1}{3} k N^3 
\ee
The physical NMSSM Higgs sector consists of 
three scalars $h^0,H^0_{1,2}$ and 
two pseudoscalars $A_{1,2}^0$.
As in the  minimal model, $\tan \beta =  v_u/v_d$ denotes the ratio
of Higgs doublet vevs $v_u=<H_u^0>=v \sin \beta$ and
$v_d=<H_d^0>=v \cos \beta$, where $v=\sqrt{2} m_W/g \simeq 174$ GeV.
The Higgs potential 
\be
V_{higgs}=V_{soft}+V_F+V_D
\ee
where
\bea
V_{soft}\!  &=&\!  m_{H_d}^2 |H_d|^2 +m_{H_u}^2 |H_u|^2 +m_{N}^2 |N|^2
-(\lambda A_\lambda H_d H_u N+ h.c.)\! -(\frac{1}{3} k A_k N^3 \! + h.c.) \\ 
V_F \!  &=&  \! |\lambda|^2 \left( |H_d|^2+|H_u|^2\right) |N|^2 + |\lambda H_d
H_u -k N^2|^2 \\
V_D \! & = & \! \frac{g^2+g^{\prime 2}}{8} \left(
|H_d|^2-|H_u|^2\right) +\frac{g^2}{2} |H_u^\dagger H_d|^2
\eea
has a global $U(1)_R$ symmetry in the limit of vanishing soft terms 
$A_k,A_\lambda \to 0$
\cite{Dobrescu:2000yn}.
If this  symmetry is  broken only
slightly, the model  naturally contains a light pseudoscalar.
Its  mass is given as
\be
m_{A_1^0}^2= 3 k x A_k +{\cal{O}}(\frac{1}{\tan \beta})
\ee
where $x=<N>$ denotes the vev of the singlet.
Note that a small $A_k$ remains small under renormalization group running
and thus protects  $m_{A_1^0}$. 

Lower bounds on CP-odd scalar masses
are not very stringent and can be
as low as $\sim 100$ MeV \cite{Dobrescu:1999gv}.
Since the coupling $h^0 A_{1}^0 A_1^0$ is not suppressed the scalar Higgs 
predominantly decays into the lighter pseudoscalars. This has
important consequences for the Tevatron and LHC Higgs searches 
\cite{Dobrescu:2000yn,Dobrescu:2000jt}.

The motivation for this work is to find out how and to what extend
the NMSSM would signal itself in rare $b$-decays and at the same time,
whether existing data provide already bounds on
the NMSSM parameter space. We employ the large
 $\tan \beta \gsim 30 $ and small $A_k \ll m_W,x$
limit and no flavor or CP violation other than in the CKM
matrix (``minimal flavor violation'').
Since a small $A_\lambda$ is not stable under radiative corrections,
we do not expand in small $A_\lambda$ and keep it finite.
Our study is based on mostly generic features of the NMSSM.
Specific analyses of the NMSSM particle spectrum
and parameter space have been carried out in a
GUT framework \cite{Ellwanger:1996gw} 
at large $\tan \beta $ \cite{Ananthanarayan:1995xq},
with gauge mediated SUSY breaking \cite{Han:1999jc} and with 
anomaly mediation \cite{Kitano:2004zd}.
For Higgs production in rare $b$-decays in other models, see 
e.g.~\cite{Frere:1981cc,Haber:1987ua}.

This paper is organized as follows:
In Section \ref{sec:amplitude} we calculate the amplitude for
$b \to s A_1^0$ decays at large $\tan \beta$.
We discuss the NMSSM parameter space in Section \ref{sec:space}.
Phenomenological bounds from FCNC decays, $B - \bar B$ mixing 
and $\Upsilon$-decays are worked out in Section \ref{sec:bounds}.
In Section \ref{sec:impact} we investigate the impact on semileptonic
and radiative rare $b$-decays. 
We also analyse how much the MSSM tree level relation
Eq.~(\ref{eq:tree}) is broken by loop corrections.
We conclude in Section \ref{sec:conclusions}.
Feynman rules and the NMSSM particle spectrum at large
$\tan \beta$ and auxiliary functions are given in Appendix \ref{app:spectrum}
and \ref{app:conventions}. In Appendix \ref{app:decayrates} we give 
decay rates of the $A_1^0$ and $b$-decay branching ratios.

\section{The $b \to s A_1^0$ amplitude at large $\tan \beta$}
\label{sec:amplitude}

The amplitude for a FCNC $b \to s$ transition into the lightest CP-odd
scalar  $A_1^0$ in the NMSSM is induced at one-loop.
In the large $\tan \beta$ limit, only two diagrams remain to be
calculated, which are shown in Figure \ref{fig:loops}. 
(We neglect the strange quark mass).
Feynman rules are given in Appendix \ref{app:rules}, see also
\cite{Rosiek:1995kg} for the MSSM and 
\cite{Franke:1995tc} for the NMSSM.

The stop chargino wave function correction is identical to 
the corresponding one in the MSSM.
Since the coupling of the $A_1^0$ to down-type fermions is
order $(\tan \beta)^0$, the 1PR diagram contributes to the $b \to s A_1^0$
amplitude at order $\tan \beta$.
The vertex correction shown in
Figure \ref{fig:loops} is the only 1PI diagram linear
in $\tan \beta$ because
{\it i} the $H^\pm W^\mp A_1^0$ coupling is $1/\tan \beta$ suppressed
since the $A_1^0$ is predominantly the gauge singlet
(the $H^+ H^- A_1^0$, $W^+ W^- A_1^0$ vertices are forbidden by CP),
{\it ii} the coupling of the $A_1^0$ to up-type quarks is $1/\tan^2 \beta$, 
{\it iii} the coupling of the $A_1^0$ to up-type squarks is $1/\tan
\beta$ which can be seen from the F-term contribution $|\partial
W/\partial H_u|^2$
and {\it iv} the only $\tan \beta$ enhancement comes from the
$b_R \tilde t_L \tilde H_d$ or $b_R t_L H_d$ vertices.

\begin{figure}[htb]
\vskip -1.2truein
\begin{center}
\includegraphics[height=7.6in,width=7in,angle=0]{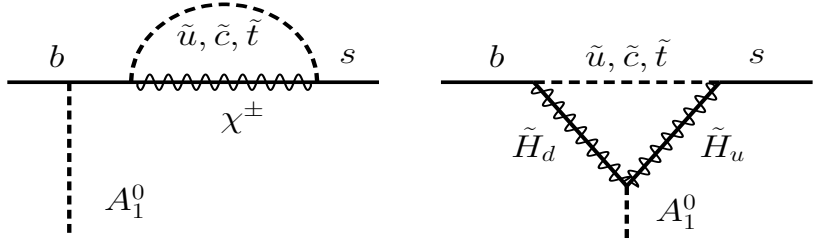}
\vskip -4.8truein
\end{center}
\caption[]{ \it The leading $b \to s A_1^0$ 
diagrams at large $\tan \beta$ in the NMSSM.}
\label{fig:loops}
\end{figure}

We obtain the following amplitude
\be
\label{eq:amplitude}
i {\cal{A}}(b \to s A_1^0)=-4 \frac{G_F}{\sqrt{2}} V_{tb} V_{ts}^* C_A 
\frac{1}{16 \pi^2} \bar s_L b_R A_1^0
\ee
where
\be
C_A=-i \frac{\tan \beta m_b}{\sqrt{2}} \left[ -\frac{\delta_-}{x} 
\sum_{i=1,2} X_i+
\lambda m_t \sin \theta_{\tilde t} \cos \theta_{\tilde t}
\sum_{l,j=1,2} 
Y_{lj} \right]
\label{eq:ca}
\ee
and $\delta_-$ parametrizes the $A_1^0 \bar b b$ coupling, see
Eq.~(\ref{eq:delta}). 
The $X,Y$ terms in Eq.~(\ref{eq:ca}) result from the wave function and
vertex correction, respectively . 
They are written as
\bea
X_i  & = &  m_{\chi_i} U_{i2} \left[ \sqrt{2} m_W
V_{i1} \left( 
-D_3(y_{c,i})+D_3(y_{1,i}) \cos^2 \theta_{\tilde t}+ D_3 (y_{2,i})
\sin^2 \theta_{\tilde t} \right) \nonumber \right. \\
&-& \left.  m_t V_{i2}
\sin \theta_{\tilde t} \cos \theta_{\tilde t} 
\left( D_3(y_{1i})-D_3(y_{2i}) \right)
\right] \\
Y_{lj}  &=&  V_{j2} U_{l2} \left[  \left( y_{1j} U_{j2} V_{l2}
-\frac{m_{\chi_l}}{m_{\chi_j}} U_{l2} V_{j2} \right) 
D_2(y_{1j},z_{lj}) \nonumber \right. \\
&& \left.  \hspace{1cm}
- \left( y_{2j} U_{j2} V_{l2} 
-\frac{m_{\chi_l}}{m_{\chi_j}} U_{l2} V_{j2} \right)  D_2(y_{2j},z_{lj})
 \right]
\eea
where
\be
y_{kj} = \frac{m_{\tilde t_k}^2}{m_{\chi_j}^2} \, ,~~~~~~
y_{cj} = \frac{m_{\tilde c}^2}{m_{\chi_j}^2} \, , ~~~~~~
z_{lj} = \frac{m_{\chi_l}^2}{m_{\chi_j}^2} 
\ee 
and $m_{\tilde t_k}, m_{\tilde c},m_{\chi_l}$ denote the stop, scharm
and chargino masses.
The stop mixing angle $\theta_{\tilde t}$, the chargino mixing matrices
$U,V$ and the loop functions $D_2,D_3$ are defined in Appendix
\ref{app:conventions}.
We used unitarity of the CKM matrix and neglected squark mixing
other than for stops and mass splitting between the first two generations. 
The $b \to d A_1^0$ amplitude is obtained by replacing
`$s$' by `$d$' everywhere in Eq.~(\ref{eq:amplitude}).
The $s \to d A_1^0$ amplitude is given correspondingly with also 
changing $m_b$ to $m_s$ in Eq.~(\ref{eq:ca}).
Note that our calculation holds for
$|\delta_\pm v/x| <\tan \beta$, see Appendix \ref{app:spectrum}. 
E.g.~for larger values of $\delta_- v/x$ the $A_1^0$
looses its mostly-singlet nature 
and more $b \to s A_1^0$ diagrams need to be calculated.

The coupling $C_A$ vanishes if the super GIM mechanism is
active, that is either all squark masses are degenerate 
or $m_{\tilde c}=m_{\tilde t_1}$ and $\theta_{\tilde t}=0$ (or $\pi$)
or $m_{\tilde c}=m_{\tilde t_2}$ and $\theta_{\tilde t}=\pi/2$.
We estimate the generic size of $C_A$ with order one stop mixing as
\be
|C_A| \simeq {\cal{O}}(\delta_- \tan \beta m_b m_t\frac{m_{\chi}}{x})
+{\cal{O}}(\lambda \tan \beta m_b m_t )
\label{eq:generic}
\ee
Since $ \lambda x$ is the NMSSM $\mu$-term which sets the mass scale
for the charginos, both terms are of comparable size.

\section{Viable points in the NMSSM parameter space \label{sec:space}}

The relevant NMSSM parameter space consists of 
$\lambda,k$ from the superpotential, the soft breaking
terms $A_\lambda, A_k$, the gaugino mass $m_2$, stop and scharm masses, 
the stop mixing angle $\theta_{\tilde t}$ and $\tan \beta$.
We evaluate all parameters at the electroweak scale.

The dimensionless couplings $\lambda$ and $k$ 
run towards smaller values, e.g.~$\lambda^2+k^2 \lsim 0.6 $ at the electroweak
scale for $\lambda,k \lsim 2 \pi$
at the high, GUT scale \cite{Miller:2003ay}. We use 
$|\lambda|, |k| \leq 1$.
Similar to the MSSM,
electroweak symmetry breaking at large $\tan \beta$ requires
\cite{Ananthanarayan:1995xq}
\be
m_{H_u}^2 = -(\lambda x)^2-\frac{m_Z^2}{2}
\ee
and therefore the product of $\lambda$ and $x$ should not exceed 
$\sim {\cal{O}}(1)$ TeV to avoid fine tuning.
On the other hand, the chargino mass scale is driven
by $\lambda x$, which should be at least ${\cal{O}}(100)$ GeV by 
experimental search limits.
We assume the singlet vev $x$ to be of the order of the Fermi scale $v$,
or at least not smaller than 100 GeV and not bigger than 3 TeV.
If $x$ exceeds this value, its
relation to the other vevs becomes unnatural and the model does not 
give a solution to the
$\mu$ problem \cite{Nilles:1982dy}. Hence, the size of $\lambda$ 
is bounded from below 
as $|\lambda| \gsim \mbox{few} \cdot 10^{-2}$
\cite{Han:1999jc,Miller:2003ay}. 
Further, the extremization condition 
\be
m_{H_d}^2 = -\lambda^2 (x^2+v^2) +m_A^2 +\frac{m_Z^2}{2}
\ee
where we defined
\be
\label{eq:ma}
m_A^2 \equiv \lambda (A_\lambda+ k x) x \tan \beta
\ee
implies some cancellation among the  $\tan \beta$ enhanced terms as 
\cite{Ananthanarayan:1995xq}
\be
\label{eq:cancel}
 A_\lambda+ k x \sim \frac{{\cal{O}}(100-1000 \, \mbox{GeV})}{\tan \beta} 
\ee
Note that $m_A$ 
sets the scale for the heavy Higgses $A_2^0,H_2^0$ and $H^ \pm$,
see Appendix \ref{app:spectrum}.

The NMSSM is further constrained by 
non-observation of Higgses and superpartners.
At large $\tan \beta$, the mass of the lightest scalar
at tree level is given as
\be
m_{h^0}^2 \simeq m_Z^2-\frac{\lambda^4 v^2}{k^2}
\ee
where we expanded Eq.~(\ref{eq:higgsmass}) in $m_Z^2/4 k^2 x^2 \ll 1$
and $\lambda^2 v/k^2 x \ll 1$. In this approximation also
$m_{H_1^0}^2 \simeq 4 k^2 x^2+\frac{\lambda^4 v^2}{k^2}$ and
the scalar mixing angle $\theta$ is small.
Like in the MSSM, the $h^0$ tree level mass cannot be bigger than the 
$Z$-mass because the raising of its upper bound in the NMSSM is 
suppressed by large $\tan \beta$.
To be phenomenologically viable,
$m_{h^0}$ has to be lifted by radiative corrections above the 
current search limit as in the MSSM \cite{Ellwanger:1993hn}.
We require the scalar tree level mass to be bigger than 89 GeV, which 
favors small $\lambda$ or $\lambda/k$ less than one.
We allow for $|m_2| \leq 1$ TeV and 
check  that  the charginos are heavier than 90 GeV.
We treat the pseudoscalar masses $m_{A_1^0}$ and $m_{A_2^0}$ 
with $m_{A_2^0} \gsim 130$ GeV  
as free parameters, i.e.~adjust $A_k$ and $A_\lambda$ accordingly.
The squark masses  and stop mixing angle are effective parameters with
$m_{\tilde t_1}> 90 $ GeV and $m_{\tilde t_2},m_{\tilde c}  \sim 1 $ TeV and 
we do not relate them to fundamental parameters in the Lagrangian.

The down-type fermion-$A_1^0$ vertex is proportional to
$\delta_- v/x$, see Eq.~(\ref{eq:A1bb}).
{}From Eqs.~(\ref{eq:ma}) and
(\ref{eq:cancel}) we obtain 
\bea
\label{eq:simple}
\frac{v}{x}\delta_- &=& \frac{v}{x} 
\left[-3 \frac{ k \lambda x^2}{m_A^2} \tan \beta +1 \right] 
\simeq  \pm 3 \frac{ k v m_\chi}{m_A^2} \tan \beta
\eea
where the second equation is a good approximation for not too large 
$m_A\lsim 500$ GeV.
It then gives a lower bound on $| \delta_- v/x|$.
In particular, for
$\tan \beta=30$, $m_A \lsim 500\, (200) \, (130)$ GeV and 
the ranges of parameters given in the preceding paragraphs,
we obtain $|\delta_- v/x| \gsim 0.1 \, (1) \, (3)$.
For  larger values of $m_A$ cancellations between the two terms in
Eq.~(\ref{eq:simple}) are possible.
Note that the $\tan \beta $ factor 
is only a formal enhancement, since it is cancelled by the one in
$m_A^2$. 
We find that
$|\delta_- v/x| \leq 62 \, (16)$ for $m_A \geq 500\, (1000)$ GeV.
Note that the small $A_\lambda \ll k x$ limit with $\delta_- \simeq -2$ 
makes its hard to satisfy Eq.(\ref{eq:cancel}).

\section{Phenomenology of the light ${A_1^0}$ \label{sec:bounds}}

We work out constraints on the mass of the $A_1^0$  
in the NMSSM at large $\tan \beta$ 
from $A_1^0$ production in rare decays (Section \ref{sec:rarebounds}),
$\bar B-B$ mixing (Section \ref{sec:mixingbounds}) and $B_s \to \mu^+ \mu^-$
decays (Section \ref{sec:Bsmumu}). 
We make use of the $b \to s A_1^0$ amplitude
calculated in Section \ref{sec:amplitude}. We scan the parameter space
in the regions discussed in Section \ref{sec:space}.
All FCNC bounds can be evaded by a sufficiently tuned-in super
GIM mechanism, see Section \ref{sec:amplitude}.
To quantify this, we demand in our numerical analysis
for the mass splitting $m_{\tilde t_2}-m_{\tilde t_1} > 50$ GeV 
while varying $m_{\tilde t_1}$
and for the stop mixing
$\epsilon < \theta_{\tilde t} < \pi/2 - \epsilon $ or
$\pi/2+ \epsilon < \theta_{\tilde t} < \pi - \epsilon $
with $\epsilon=0.05$. 
Bounds from other processes are discussed in Section \ref{sec:otherbounds}.

Many experimental constraints  we use here apply only if the
$A_1^0$ is sufficiently stable, i.e.~leaves the detector as missing energy.
This happens if the pseudoscalar width is smaller than 
$E_{A}/(m_{A_1^0} \, d)$,
where $d \sim {\cal{O}}(10)$ m is the size of the detector and $E_{A}$ the
$A_1^0$ energy in the lab frame.
We work out bounds on $m_{A_1^0}$  as a function of 
$|\delta_- v/x|$. If this coupling gets
smaller, the pseudoscalar decay rate 
decreases, and a heavier Higgs will become missing energy and vice versa. 
For decay rates of the $A_1^0$, see Appendix \ref{app:decayrates}.

For Higgs masses below $2 m_\mu$ only the $e^+ e^-$ and $\gamma \gamma$ 
decay channels are relevant. 
(The $A_1^0 \to \pi^0 \gamma$ decay is forbidden by CP and angular momentum
conservation and the $A_1^0 \to \pi^0 \gamma \gamma$ decay is 
suppressed with respect to 
the dielectron mode by phase space and powers of $\alpha$).
The $\gamma \gamma $ mode can compete with $A_1^0 \to f\bar f $ decays only
near the dimuon threshold.
This weakens the missing energy bounds in that region.

\begin{figure}[htb]
\vskip 0.0truein
\begin{center}
\includegraphics[height=5.0in,width=3.9in,angle=270]{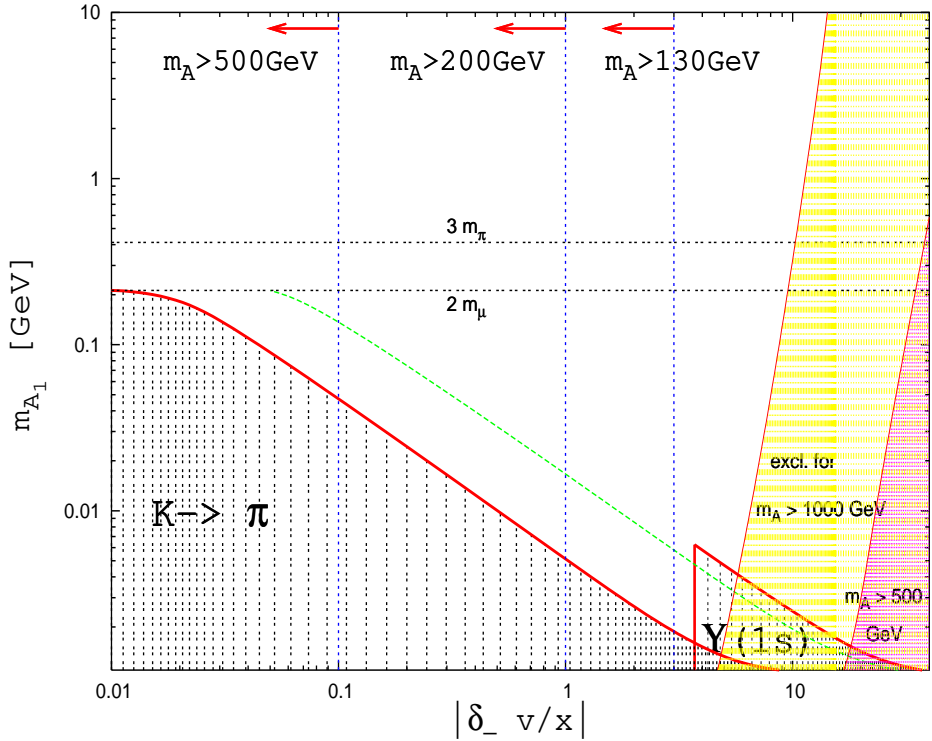}
\vskip 0.0truein
\end{center}
\caption[]{ \it Constraints on the $A_1^0$ mass as a function of 
$|\delta_- v/x|$ at $\tan \beta=30$ 
in the NMSSM. Shaded regions are excluded.  
The left bottom corner is excluded by
rare $K$-decay, see Eq.~(\ref{eq:Ktopi}). 
The triangular region to the lower right is obtained from
radiative $\Upsilon(1s)$ decays, see Eq.~(\ref{eq:yps}). 
The 
region to the left of the vertical dashed blue lines can only be reached if 
$m_A$ is bigger than the value indicated, see Section \ref{sec:space}. 
Constraints from $\Delta m_d$ are
given for $m_A \geq 500$ GeV and 
$m_A\geq 1000$ GeV. We also show the missing energy condition for 
$B \to K$ decays given in Eq.~(\ref{eq:nunubound}) (dashed green line).
The vertical dashed lines indicate $m_{A_1^0}= 2 m_\mu$ and $3 m_\pi$.}
\label{fig:bounds}
\end{figure}

The point we want to make is to show  that  in the NMSSM ${A_1^0}$ masses  
in the GeV range and below are not ruled out.
This is summarized in Figure \ref{fig:bounds}. 
For details see the following subsections.
All experminental bounds are taken at 90 \% confidence level.
The requisite $b \to A_1^0$
branching ratios are given in Appendix \ref{app:decayrates}.
We recall that our approximation breaks down if  $|\delta_\pm v/x|$ approaches
$\tan \beta$.

\subsection{Rare $K$ and $B$-decays \label{sec:rarebounds}}

If the Higgs boson is light enough,
it can be produced in $b \to s A_1^0$ or $s \to d A_1^0$ processes.
We analyze what bounds exist depending on the mass of the $A_1^0$.

\subsubsection{$2 m_e < m_{A_1^0} <2 m_\mu$ \label{sec:lessmmu}}

When produced in rare $B$-meson decays
the $A_1^0$ decays outside of the detector if
\be
m_{A_1^0} \lsim 17 \; \mbox{MeV} /| \delta_- v/x|
\label{eq:nunubound}
\ee
In this region the CLEO bound 
${\cal{B}}(B \to K_S^0 X^0) < 5.3 \cdot 10^{-5}$ \cite{Ammar:2001gi}
applies. There is a similar missing energy bound from BaBar 
${\cal{B}}(B^- \to K^- \nu \bar \nu) < 7.0 \cdot 10^{-5}$ \cite{Aubert:2003yh}
\footnote{The experimental cut on the $K$ momentum
$|\vec p_K| > 1.5$ GeV is no restriction for light $m_{A_1^0} \ll
m_B$ discussed here.}.
We find that masses in the range given in
Eq.~(\ref{eq:nunubound}) are disfavored
since the $B \to K A_1^0$ decay, see Eq.~(\ref{eq:BKA}), 
would happen too rapidly for most of the parameter space, although 
cannot rigorously be excluded.
We stress that the size of the coupling $C_A$ can be quite large, 
see Eq.~(\ref{eq:generic}) and already ${\cal{B}}(B \to X_s A_1^0) < 1$ 
cuts out a fraction of NMSSM points.

Rare decays into $e^+ e^-$ constrain  Higgs masses below the muon threshold.
However, 
the measurements of the inclusive $B \to X_s e^+ e^-$ branching ratios
contain cuts on the dilepton mass $m_{e e} \gsim 2 m_\mu$ 
\cite{Aubert:2003rv,Kaneko:2002mr}.
In the analysis of $B \to K^{(*)} e^+
e^-$ decays Belle applies  $m_{e e} > 0.14$ GeV \cite{Ishikawa:2003cp}, whereas
BaBar \cite{Aubert:2003cm} has no cut, but 
the efficiency is low in that region due to conversion photons.
Likewise, measurements of $K^+ \to \pi^+ e^+ e^-$ decays employ a high mass
trigger \cite{Appel:1999yq}.
Since also close to $2 m_\mu$ the two-photon decay of the 
$A_1^0$ becomes sizeable,
we do not take the $e^+ e^-$ data into account.

The bound ${\cal{B}}(K^+ \to \pi^+ A_1^0) < 4.5 \cdot 10^{-11}$ 
\cite{Adler:2002hy} is applicable if the $ A_1^0$ becomes
sufficiently stable to escape the detector. This happens for masses
\be
\label{eq:Ktopi}
m_{A_1^0} \lsim 5 \, \mbox{MeV}/|\delta_- v/x|
\ee 
which then are excluded. The $K$-decay  bound is five orders
of magnitude better than the one from $B \to K$ decays, 
because the CKM and mass suppression of the $K \to \pi A_1^0$ decay rate
is compensated by the difference in life time 
$|V_{td}/V_{ts}|^2 (m_K/m_B)^3 \tau(K^+)/\tau(B^+) \simeq 0.24$ 
\cite{Hagiwara:fs}, see Eq.~(\ref{eq:BKA}) and its $K \to \pi$ counterpart.

\subsubsection{$2 m_\mu <m_{A_1^0} < 2 m_\tau$ \label{sec:hadronic}}

$A_1^0$ decays into a muon pair are
included in $B \to X_s \mu^+ \mu^-$ signals. 
Comparison of the $B \to X_s A_1^0$ branching ratio, see
Eq.~(\ref{eq:bsA}), with the data 
${\cal{B}}(B \to X_s \mu^+ \mu^-)\leq 10.4 \cdot 10^{-6}$
\cite{Hiller:2003js,Aubert:2003rv,Kaneko:2002mr}
shows that this is very unlikely. 
The same happens in $K \to \pi  \mu^+ \mu^-$ decays, which
for $m_{A_1^0} < m_K-m_\pi$ can hide a pseudoscalar decaying into muons.
With
${\cal{B}}(K^+ \to \pi^+ \mu^+ \mu^-) \leq 10.4 \cdot 10^{-8}$ 
\cite{Hagiwara:fs}
only a tiny number of points survives the scan.
All allowed points are at the GIM boundary $\theta_{\tilde t} \simeq \pi/2$, 
which is set by our value of the cut-off $\epsilon$.

Above the $3 \pi$ threshold sizeable hadronic decays open up.
(The $A_1^0 \to 2 \pi  \gamma$ decay is suppressed with respect to the
dimuon channel by phase space and $\alpha$, whereas $A_1^0 \to 2 \pi$ 
decay is forbidden by CP invariance.)
For the  $A_1^0$ decaying hadronically into a strange final state we use 
${\cal{B}}(b \to s g) < 9 \%$ \cite{Coan:1997ye}.
This thins out the NMSSM model space for $3 m_\pi < m_{A_1^0} < 2 m_\tau$,
but cannot exclude this region. (We use 
${\cal{B}}(b \to s A_1^0) > {\cal{B}}(b \to d A_1^0)$.)

\subsubsection{$ 2 m_\tau < m_{A_1^0} \lsim m_B$}

If the $A_1^0$ is above the tau threshold, most of the time 
it decays into
$\tau^+ \tau^-$ because its coupling to 
$c \bar c$ is $\tan^2 \beta$ suppressed. Similar to the
constraint on the hadronically decaying pseudoscalar, 
see Section \ref{sec:hadronic}, the mildly model-dependent bound
${\cal{B}}(B \to X_s \tau^+ \tau^-) < 5$ \% \cite{Grossman:1996qj}
is not a challenge to the light CP-odd Higgs scenario.

\subsection{NMSSM neutral Higgs contributions to $B -\bar B$ mixing
\label{sec:mixingbounds}}

We calculate the contribution to $B -\bar B$ mixing from
pseudoscalar $A_{1,2}^0$ and scalar $h^0,H_{1,2}^0$ Higgs exchange 
in the NMSSM at large $\tan \beta$.
It arises at two-loop from double insertion of the 
FCNC $\bar s b$-Higgs vertices such as generated by the 
diagrams in Figure \ref{fig:loops} for the $A_1^0$
and an intermediate boson propagator. 
The dominant diagrams induced by the heavy Higgses, i.e.~the 
ones other than the lightest CP-odd 
scalar are the wave function corrections contributions with
$A_2^0,H_2^0$ exchange, see the Feynman rules in Appendix \ref{app:rules}.
They can compete with one-loop contributions 
such as the Standard Model (SM) box diagrams due to 
their $\tan^4 \beta$ enhancement.
Contributions from $h^0,H_1^0$ are subleading in $\tan \beta$.
We use an effective Hamiltonian ($q=d,s$)
\be
{\cal{H}}_{eff}^{\Delta B =2}= \frac{G_F^2 m_W^2}{16 \pi^2} 
(V_{tb} {V_{tq}^*})^2 \sum_i C_i Q_i
\ee
where some of the relevant 
operators are written as, see e.g.~\cite{Buras:2002vd}
\bea
Q^{VLL}  & = & (\bar q_L \gamma_\mu b_L) (\bar q_L \gamma^\mu b_L) \\
Q_1^{SRR} & =& (\bar q_L b_R) (\bar q_L b_R) \\
Q_1^{SLR} & =& (\bar q_R b_L) (\bar q_L b_R) 
\eea
The SM contribution is in the coefficient $C^{VLL}$.
The $A_2^0,H_2^0$ masses are degenerate at large $\tan \beta$
and their respective contributions to $Q_1^{SRR}$ cancel each other 
just like in the MSSM, see Eq.~(\ref{eq:tree}).
They do, however, contribute to the operator $Q_1^{SLR}$ at order
$m_q/m_b$, and are important for $B_s$-mesons.
(This is the famous double penguin (DP) contribution of
the MSSM \cite{Isidori:2001fv,Buras:2002vd}.)

We obtain at order $\tan^2 \beta/m_{A_1^0}^2$ in the NMSSM from 
$A_1^0$ boson exchange
\be
C_1^{SRR}(\mu_t)=-\frac{1}{4 \pi^2} \frac{C_A^2}{m_W^2 m_{A_1^0}^2}
\ee
at the high, electroweak (matching) scale $\mu_t$.
Finite widths effects are neglected.
We define the size of the $B_q -\bar B_q$ mass difference $\Delta m_q$
with respect to its SM value as
\be
\frac{\Delta m_q}{(\Delta m_q)_{SM}} = 1+f_q
\ee
where
\be
\label{eq:fq}
f_q=\frac{\bar P_1^{SLL} }{S_0(\mu_t)} C_1^{SRR}(\mu_t) 
\ee
and $S_0(\mu_t) = 2.38$ and $\bar P_1^{SLL}=-0.37$ \cite{Buras:2002vd}.
In Eq.~(\ref{eq:fq}) the 
NMSSM contribution to $\Delta m_d$ by neutral Higgs exchange
in the $m_q=0$ limit has been given.
To be in agreement with data we require 
$f_d > -0.6$ ($f_d$ is negative).
This includes $20$ \% uncertainty 
and allows for cancellations between the $A_1^0$ contribution and 
the charged Higgs, chargino boxes and the double penguins. 
We assume similar sizes as in the MSSM, where
$-0.2 \lsim f_d^{H^\pm}+f_d^{\chi^\pm}+f_d^{DP} \lsim 0.4 $ 
\cite{Buras:2002vd}.
We find constraints for larger values of $|\delta_- v/x|$ and 
$m_A \geq 500$ GeV, which are
displayed in Figure \ref{fig:bounds} for $\tan \beta=30$. 
The other branch with $1+f_d <0$, where the NMSSM correction
is larger than the SM box gives very similar constraints and is not shown.
The leading $A_1^0$ contribution to $B -\bar B$ mixing is universal
in minimal flavor violation, $f_d =f_s$, since we neglect light quark masses.

\subsection{$B_s \to \mu^+ \mu^-$ decays \label{sec:Bsmumu}}

We work out the contributions to $B_s \to \mu^+ \mu^-$ decays
from neutral Higgs exchanges in the large $\tan \beta$ limit of the NMSSM.
With the effective Hamiltonian
\be
\label{eq:Heff}
{\cal{H}}_{eff}= -\frac{G_F}{\sqrt{2}} 
V_{tb} V_{ts}^* \sum_i C_i {\cal{O}}_i
\ee
where
\be
{\cal{O}}_S = \frac{e^2}{16 \pi^2} \bar s_L b_R \bar \ell \ell \, , ~~~~~~
{\cal{O}}_P = \frac{e^2}{16 \pi^2} \bar s_L b_R \bar \ell \gamma_5 \ell
\ee
we obtain at the electroweak  scale
(in parentheses is given the particle
that induces a particular Wilson coefficient)
\be
C_S=C_S(H_2^0)= -C_P(A_2^0) \, , ~~~~~~ C_P= C_P(A_1^0)+C_P(A_2^0)
\ee
where
\bea
C_P(A_1^0)& =& m_b m_\ell \tan \beta \frac{v}{4 m_W^2 \sin^2 \theta_W} 
\left( \frac{v \delta_- }{x} \right) \frac{1}{m_{B_s}^2-m_{A_1^0}^2} 
\nonumber \\
&\times  &\left[ -\frac{\delta_-}{x} 
\sum_{i=1,2} X_i+
\lambda m_t \sin \theta_{\tilde t} \cos \theta_{\tilde t}
\sum_{l,j=1,2} 
Y_{lj} \right]
\\
C_P(A_2^0)& =& m_b m_\ell \tan^3 \beta \frac{1}{4 m_A^2 m_W^2 \sin^2 \theta_W} 
\sum_{i=1,2} X_i
\eea
The expressions for $X$ and $Y$ are given in Section \ref{sec:amplitude}.
Our result for the $A^0_2,H^0_2$ contributions agrees with the 
corresponding MSSM calculations \cite{Bobeth:2001sq}. 
Note that the contributions from $A^0_2$ and $H^0_2$ 
are equal with opposite sign. Similar to
$B-\bar  B$ mixing discussed in Section \ref{sec:mixingbounds}, 
the scalars $h^0$ and $H_1^0$ contribute at subleading order in $\tan \beta$.

The coefficients $C_{S,P}$ are model-independently constrained by data on the 
$B_s \to \mu^+ \mu^-$ branching ratio. With 
${\cal{B}}(B_s \to \mu^+ \mu^-) <5.8 \cdot 10^{-7}$ \cite{Acosta:2004xj}
we obtain at the scale $\mu=m_W$
\bea\label{eq:CSPbound}
\sqrt{|C_S|^2+|C_P+\delta_{10}|^2 } \leq 1.3
\Bigg[\frac{{\mathcal{B}}( B_s \to \mu^+ \mu^-)}{5.8 \times
10^{-7}}\Bigg]^{1/2} 
\Bigg[\frac{238\, \mbox{MeV}}{f_{B_s}}\Bigg] 
\eea
Here, $\delta_{10}$ stems from the operator 
${\cal{O}}_{10} \propto \bar s_L \gamma_\mu b_L \bar \ell \gamma^\mu \gamma_5 \ell$, see \cite{Hiller:2003js} for details. 
We find with
$\delta_{10}^{SM} =-0.095$ 
\footnote{Susy effects are not $\tan \beta$ enhanced in 
${\cal{O}}_{10}$
and are small with minimal flavor violation \cite{Ali:1999mm}.} 
at $\tan \beta=30$ the upper limits
$|\delta_- v/x| \lsim 42 \, (16) $ for $m_A =500 \, (1000)$ GeV,
which are
are weaker than the corresponding $\Delta m_d$ ones. 
The expressions for the CKM suppressed 
$B_d \to \mu^+ \mu^-$ decay are readily obtained. 
Its experimental constraint is not as good as the $B_s$ one,
but we can cut out both
$m_{A_1^0} \simeq m_{B_d}$ and $m_{A_1^0} \simeq m_{B_s}$.

\subsection{Non-FCNC bounds \label{sec:otherbounds}}

The bounds from radiative $\Upsilon$-decays
apply if the $ A_1^0$ leaves the detector unseen 
\cite{Hagiwara:fs,Balest:1994ch}.
Due to the larger boost the critical width to do so is
larger than in $B$-meson decays by $m_{\Upsilon}/m_B$.
We use 
${\cal{B}}( \Upsilon(1s) \to  A^0 \gamma ) < 1.3 \cdot 10^{-5} $ 
\cite{Balest:1994ch} and obtain with Eq.~(\ref{eq:upsbr})
\be
\label{eq:yps}
|\delta_- v/x | \lsim 3.7 ~~~~~ \mbox{ for} ~~~ m_{A_1^0} \lsim 23 \, \mbox{MeV}/|\delta_- v/x| < 2 m_\mu
\ee
Furthermore, we get an upper bound $|\delta_- v/x| \lsim 100$ from
${\cal{B}}(\Upsilon(1s) \to  A_1^0 \gamma) <1$.

Mass bounds from hadronic collisions are 
not better than few to 200 MeV and 
astro physics gives $m_{A_1^0} \gsim 0.2 $ MeV \cite{Hagiwara:fs}, 
which contain some  model dependence.

\section{Implications for $b \to s \ell^+ \ell^-$ and $ b \to s
\gamma, g$}
\label{sec:impact}

Similar to the operators ${\mathcal{O}}_{S,P}$ discussed in 
Section \ref{sec:Bsmumu}, 
the NMSSM Higgs sector also induces 
contributions to 4-Fermi operators with quarks and leptons 
($f$ denotes a fermion)
\begin{eqnarray}\label{new:ops:scalar}
{\cal{O}}_{L}^f =  \bar{s}_L b_R \bar{f}_R f_L \, , \quad
{\cal{O}}_{R}^f= \bar{s}_L  b_R \bar{f}_L f_R
\end{eqnarray}
where
\be
C_{L, R}^f = \frac{e^2}{16 \pi^2} \frac{m_f}{m_\mu} (C_S \mp C_P)
\ee
These couplings arise in the NMSSM at large $\tan \beta$, where
\be
C_S-C_P=-2 C_P(A_2^0)-C_P(A_1^0) \, , ~~~~~~  C_S+C_P=C_P(A_1^0)
\ee
This is different from the MSSM, where the $A_1^0$ contribution is absent and
the sum of $C_S$ and $C_P$ and hence $C_R^f$ vanish.
We discuss corrections 
to this tree level statement in Section \ref{sec:CRsize}.
All Wilson coefficients refer to the Hamiltonian
in Eq.~(\ref{eq:Heff}) and are evaluated at the scale $\mu=m_W$
unless otherwise stated.

The constraint given in
Eq.~(\ref{eq:CSPbound}) implies for the Wilson coefficients for $b$-quarks
(we update the findings of 
Ref.~\cite{Hiller:2003js} with the improved $B_s \to \mu^+ \mu^-$ bound 
\cite{Acosta:2004xj}) 
\begin{eqnarray}
\label{eq:clbbound}
\sqrt{ |C_{L}^b|^2+|C_{R}^b|^2 } \lsim 0.03 
\end{eqnarray}
The operators ${\mathcal{O}}_{L,R}^b$
enter radiative and semileptonic rare $b \to s $ decays at one-loop
 \cite{Hiller:2003js,Borzumati:1999qt}.
With the bound in Eq.~(\ref{eq:clbbound}) the new physics effect 
from ${\mathcal{O}}_{L}^b$ is small, at the percent level 
\cite{Hiller:2003js}. 
However, the renormalization effect induced at leading log by 
${\cal{O}}_R^b$ can be large for the
photon and gluon dipole operators ${\cal{O}}_{7}$ and ${\cal{O}}_{8}$, 
which can be written as
\be
{\cal{O}}_{7 } = \frac{e}{16\pi^2} m_b
\bar{s}_L \sigma_{\mu \nu}  b_R F^{\mu \nu}, \quad 
{\cal{O}}_{8 }=\frac{g_s}{16\pi^2} m_b
\bar{s}_{L \alpha} \sigma_{\mu \nu} T^a_{\alpha \beta} 
b_{R \beta}G^{a \mu \nu}
\ee 
To be specific, 
we normalize their coefficients 
to the ones in the SM, 
and denote this ratio by  $\xi$,  such that $\xi^{SM}=1$. With 
(see \cite{Hiller:2003js} for details)
\bea
\xi_7(m_b)& = & 0.514 + 0.450\, \xi_7(m_W) + 0.035\, \xi_8(m_W)  -  
2.319 \,C_R^b \, , \\
\xi_8(m_b)&=&  0.542 + 0.458\, \xi_8(m_W)  +19.790\,C_R^b  
\eea
and Eq.~(\ref{eq:clbbound})
corrections of up to $7 \% $  and $59 \% $ to $\xi_7$ and $\xi_8$
are possible. This has impact on the extraction of Wilson 
coefficients in $b \to s \gamma$, $b \to s g$ and 
$b \to s \ell^+ \ell^-$ decays \cite{Hiller:2003js}.
For a full analysis of  these decays, also the matching contributions
to $C_{7,8}$ from neutral Higgs loops in the large $\tan \beta$ NMSSM
have to be calculated.
Note that $\tan \beta$ enhanced corrections to the $b$-quark
mass, CKM elements and FCNCs from 
non-holomorphic terms 
arise \cite{Babu:1999hn,Isidori:2001fv,Buras:2002vd}. 
We leave this for future work.

\subsection{Estimates of $C_S+C_P$ and $C_R^b$ \label{sec:CRsize}}

\begin{figure}[htb]
\vskip 0.0truein
\begin{center}
\includegraphics[height=5.0in,width=3.9in,angle=270]{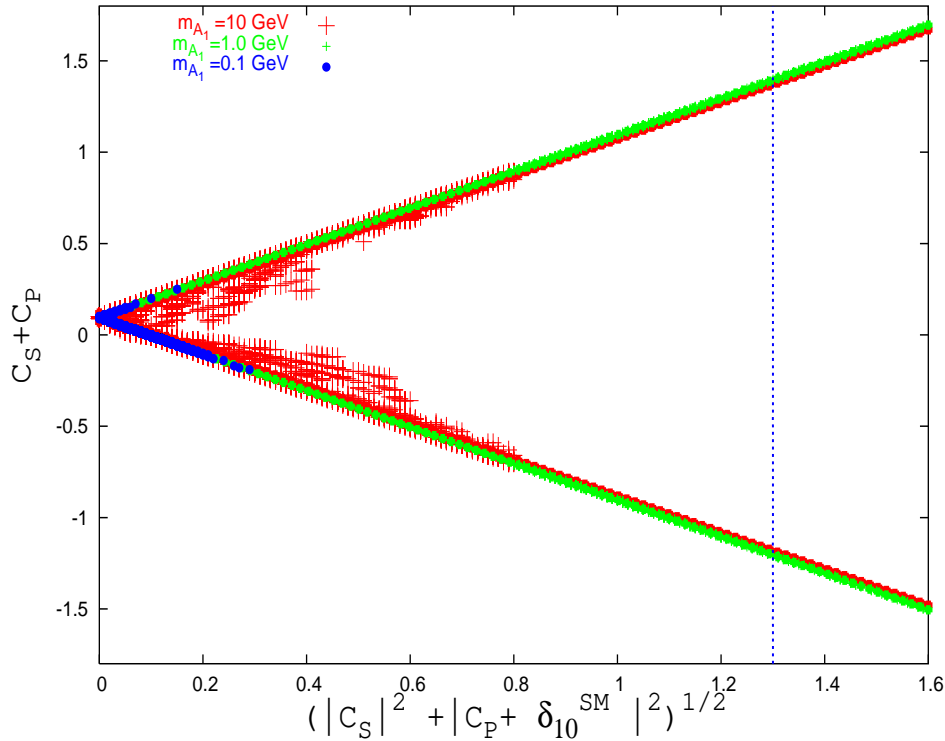}
\vskip 0.0truein
\end{center}
\caption[]{ \it The correlation between $C_S+C_P$ 
and $\sqrt{|C_S|^2+|C_P+\delta_{10}^{SM}|^2 }$ 
in the NMSSM for $\tan \beta=30$, $m_{A}=500$ GeV and 
$m_{A_1^0}=0.1 , 1 $ and $10$ GeV. Also shown is the experimental upper
bound given in Eq.~(\ref{eq:CSPbound}) (dashed line).}
\label{fig:Cr}
\end{figure}

We work out the NMSSM reach in $C_S+C_P$ by taking into account
all constraints discussed in the previous Sections \ref{sec:space} and 
\ref{sec:bounds}.
The value of $C_S+C_P$ can saturate its upper bound given in 
Eq.~(\ref{eq:CSPbound}) for large ranges of the parameter space.
If the $A_1^0$ gets very light, however, the $b \to s A_1^0$ coupling 
$C_A$ has to decrease
and  $C_S+C_P$ is small, e.g.~for $m_{A_1^0}=10$ MeV is
$|C_S+C_P|\leq 0.06$.
For intermediate masses the $A_1^0$ contribution dominates over the
one from the heavy pseudoscalar, that is $|C_P| \gg C_S$ and
$|C_R^b| \simeq |C_L^b| \lsim 0.024$.
This is illustrated in Figure \ref{fig:Cr}, where
we show $C_S+C_P$ as a function of 
$\sqrt{|C_S|^2+|C_P+\delta_{10}^{SM}|^2} \propto 
\sqrt{{\cal{B}}(B_s \to \mu^+ \mu^-)}$ , see Eq.~(\ref{eq:CSPbound}), for
$m_A=500$ GeV and different values of $m_{A_1^0}$.

In the MSSM the size of $C_S+C_P$ is driven by the relation 
Eq.~(\ref{eq:tree}), which is not protected from 
radiative corrections.
To study their size we employ the two-loop calculation encoded in 
{\it FeynHiggs }v.~2.02 \cite{feynhiggshome}.
By scanning the MSSM parameter space
we find 
\be
| \frac{C_S+C_P}{C_S-C_P} |_{\mbox{\scriptsize \it MSSM}} < 0.2 
~~~~\mbox{or}~~~~|C_S+C_P|_{\mbox{\scriptsize \it MSSM}}<0.08
 ~~~~\mbox{and}~~~~|C_R^b|_{\mbox{\scriptsize \it MSSM}}<1.3 \cdot 10^{-3}
\ee
The smallness of $C_S+C_P$ is a feature of the
Higgs sector of the MSSM. It holds also with
flavor violation beyond the CKM matrix.
As a result, the logarithmic renormalization of the dipole operators from
neutral (pseudo)scalars is tiny in this model.
For example, consider additional
right handed currents, which induce contributions to the helicity 
flipped operators ${\cal{O}}^\prime_i$, i.e.~the ones obtained from 
${\cal{O}}_i$ with right $R$ and 
left $L$ chiralities interchanged. In this case,
$C_S^\prime-C_P^\prime$
mixes onto the flipped dipole operators ${\cal{O}}^\prime_{7,8}$, but
$C_S^\prime=C_P^\prime$ in the large $\tan \beta $ MSSM 
\cite{Isidori:2001fv}.

\section{Conclusions \label{sec:conclusions}}

We investigated the phenomenology of the light pseudoscalar $A_1^0$ which 
lives  in the NMSSM spectrum at large $\tan \beta$. 
The $A_1^0$ has suppressed 
gauge interactions but couples to Higgses and down-type matter.
We calculated the $b \to s A_1^0$ amplitude at leading order in $\tan \beta$.
Based on this, we estimated the NMSSM contributions to rare $K$, $B$ and 
radiative $\Upsilon$-decays with the $A_1^0$ in the final state, 
$B_s \to \mu^+ \mu^-$ decays and $B -\bar B$ mixing.
We showed that
low energy data provide constraints on the $A_1^0$ mass and couplings, 
but leave masses down to 
${\cal{O}}(10)$ MeV viable, see Figure \ref{fig:bounds}.
The $A_1^0$ predominantly decays into $\tau^+ \tau^-$ for 
$2 m_\tau < m_{A_1^0} < 2 m_b$,
light hadrons for $3 m_\pi < m_{A_1^0} < 2 m_\tau$
and $e^+ e^-$or $\gamma \gamma$ for  
$2 m_e < m_{A_1^0} < 2 m_\mu$. In the latter range, the $A_1^0$ can live
long enough to leave detectors undecayed, depending on $\delta_- v/x$.
For masses within  $2 m_\mu < m_{A_1^0} < 3 m_\pi $ the $A_1^0$ 
decays mostly into muon pairs. Like the one from $B \to K$ decays
given in Eq.~(\ref{eq:nunubound}), this mass range has very
tight FCNC constraints, see Section \ref{sec:hadronic}, but is not 
ruled out.

The $A_1^0$ can be searched for with improved measurements of
$\Upsilon$-decays or $B \to K$ plus missing energy.
The latter needs a high $K$-momentum cut 
to suppress the $B \to K \nu \bar \nu$ background.
For $ m_{A_1^0}$ above the $\Psi^\prime$ mass the pseudoscalar can be seen 
in $b \to s \tau^+ \tau^-$ decays. The required sensitivity for e.g.~the
$B \to X_s \tau^+ \tau^-$ branching ratio is 
${\cal{B}}(B \to X_s \mu^+ \mu^-) m_\tau^2/m_\mu^2 \lsim 10^{-3}$.

The NMSSM has different implications for $b$-physics than the MSSM.
In particular, the leading log neutral Higgs contribution 
to radiative $b \to s \gamma$ and $b \to s g$ decays
is tiny in the latter, but can reach
experimental upper limits in the former, see Section \ref{sec:CRsize}.
Furthermore, the MSSM correlation between the $B_s \to \mu^+  \mu^-$ 
branching ratio and 
$B_s -\bar B_s$ mixing \cite{Buras:2002wq} breaks down due to the
additional pseudoscalar.
For example, for  small $|C_S/C_P|$ the lighter CP-odd Higgs dominates the 
$B_s \to \mu^+  \mu^-$ rate, which can be anything up to the 
experimental bound, see Figure \ref{fig:Cr}. At the same time $\Delta m_s$ 
is near its SM value because the leading $A_1^0$ contribution is independent of the light quark flavor and constrained by $\Delta m_d$, and the 
double penguin from $A_2^0$ is suppressed.
This is in contrast to the MSSM, where a SM-like
$\Delta m_s$ implies an upper bound on 
${\cal{B}}(B_s \to \mu^+ \mu^-)$.

\bigskip

\noindent
{\bf Acknowledgements}
G.H.~would like to thank Gerhard Buchalla, Yuval Grossman, Howie Haber and 
Anders Ryd for useful comments, Sven Heinemeyer for
{\it FeynHiggs} support and Bogdan Dobrescu for collaboration at an 
early stage of this project. G.H.~gratefully acknowledges the hospitality 
of the SLAC theory group.

\begin{appendix}
\renewcommand{\theequation}{\Alph{section}-\arabic{equation}}

\setcounter{equation}{0}
\section{Higgs spectrum and couplings \label{app:spectrum}}

We give the tree level Higgs spectrum, mixing angles in the
minimal flavor and CP violating 
NMSSM at large $\tan \beta$ and $A_k \ll m_W,x$.
The mass matrices in gauge eigenstates can be seen in
\cite{Ellis:1988er}.

The mass eigenstates of the pseudoscalar mixing matrix can be written as
\bea
\left( \begin{array}{c} A_1^0 \\ A_2^0 \end{array} \right)= 
\left( \begin{array}{cc}
     \cos \gamma & \sin \gamma \\
     -\sin \gamma & \cos \gamma  \end{array} \right)
\left( \begin{array}{c} A^0 \\ N_I \end{array} \right)
\eea
where $N_I=\rm Im N/\sqrt{2}$, 
$A^0=\sqrt{2} (\sin \beta \, \rm Im \, H_d^0 +\cos \beta \, \rm Im \,
H_u^0)$ and the Goldstone boson is given as 
$G^0=\sqrt{2} (\cos \beta \, \rm Im \, H_d^0 -\sin \beta \, \rm Im \, H_u^0)$.
The mixing angle and masses read as
\be
\label{eq:gamma}
\gamma =  \frac{\pi}{2}+ \frac{v}{x \tan \beta} \delta_- +
{\cal{O}}(\frac{1}{\tan^2 \beta}) ~~~\mbox{i.e.}
~~\sin \gamma\simeq 1 \, ,~~ 
\cos \gamma \simeq -\frac{v}{x \tan \beta} \delta_-
\ee
and 
\be
m_{A_1^0}^2  =3 k x A_k  \, ,~~~~~~~ 
m_{A_2^0}^2 = m_A^2 
\ee
where we defined
\be
\label{eq:delta}
\delta_\mp = \frac{A_\lambda \mp 2 k x}{A_\lambda + k x}
\ee
and $m_A^2$ is given in Eq.~(\ref{eq:ma}).

The scalar mass matrix  
can be diagonalized analytically in the large $\tan \beta $ limit 
by first decoupling the heaviest 
state and then rotating the remaining 2 by 2 block by the angle $\theta$
along the lines of Ref.~\cite{Miller:2003ay}. The result can be written as
\bea
\left( \begin{array}{c} h^0 \\ H_1^0 \\ H_2^0 \end{array} \right)= 
{\sqrt{2}} \left( \begin{array}{ccc}
 \frac{1}{\tan \beta} (\cos \theta -\frac{ v}{x} \delta_+ \sin \theta)   \; \;
& \cos \theta  \; & - \sin \theta  \\
\frac{1}{\tan \beta} (\sin \theta +\frac{v}{x} \delta_+
\cos \theta) \;  \;
    &  \sin \theta  \; & \cos \theta \\
    1 \; \;  & \frac{-1}{\tan \beta} \; & \frac{-v}{x \tan \beta}
\delta_+  \end{array} \right)
\left( \begin{array}{c} \rm Re \, H_d^0 -v_d\\ \rm Re \, H_u^0 -v_u\\ \rm Re
\, N -x \end{array} \right)
\eea
with the mixing angle and scalar masses 
\bea
\tan 2 \theta & = &  \frac{4 \lambda^2 v x}{ 4 k^2 x^2-m_Z^2} \\
m_{H_2^0}^2 &=&m_A^2 \, ,~~~~
m_{h^0,H_1^0}^2 = \frac{1}{2} \left[ 4 k^2 x^2 + m_Z^2 \mp \sqrt{(4
k^2 x^2 -m_Z^2)^2 + 16 \lambda^4 x^2 v^2 }
\right]
\label{eq:higgsmass}
\eea
The mass of the charged Higgs is given as
\be
m^2_{H^\pm}=m_A^2+m_W^2- \lambda^2 v^2
\ee

\subsection{Feynman rules \label{app:rules}}

Feynman rules can be read off the Lagrangians given at 
leading order in $\tan \beta$.
Note that $A_2^0 \simeq -A^0_{MSSM}$ in this limit.

Couplings to up (u) and down (d) type fermions 
\bea
\label{eq:A1bb}
 {\cal{L}}_{A^0_i \bar d d}&=& -i \frac{g m_d}{2 m_W}  \left(
\frac{v}{x} \delta_- A^0_1, \tan \beta A^0_2 \right) \bar d \gamma_5 d \\
 {\cal{L}}_{A^0_i \bar u u}&=&
-i \frac{g m_u}{2 m_W}  
\frac{1}{\tan \beta}
\left(\frac{ v }{x \tan \beta} \delta_- A^0_1 , A^0_2 \right) 
\bar u \gamma_5 u \\
 {\cal{L}}_{(h^0,H^0_i) \bar d d}&=&  -\frac{g m_d}{2 m_W}  
\left( ( \cos \theta -\frac{ v}{x} \delta_+ \sin \theta) h^0,  
\nonumber \right. \\
&\mbox{}& \left. (\sin \theta
+\frac{v}{x} \delta_+ \cos \theta) H^0_1, \tan \beta H^0_2 \right) \bar d  d
\\
 {\cal{L}}_{(h^0,H^0_i) \bar u u}&=&  -\frac{g m_u}{2 m_W}  
\left( \cos \theta h^0, \sin \theta H^0_1,
- \frac{H_2^0 }{\tan \beta} \right) \bar u u 
\eea
Couplings to charginos
\be
 {\cal{L}}_{A^0_1  \chi^+ \chi^- }= 
+i \frac{\lambda}{\sqrt{2}} A_1^0 \bar \chi^+_i \left[ U_{i2} V_{j2} L-
U_{j2} V_{i2} R \right] \chi^+_j
\ee
where $L,R=(1 \mp \gamma_5)/2$ are chiral projectors.

\renewcommand{\theequation}{\Alph{section}-\arabic{equation}}
\setcounter{equation}{0}
\section{Conventions, loop functions \label{app:conventions}}

The chargino mass matrix is written as ($\mu_{MSSM}=-\lambda x $)
\bea
M_{\chi^\pm}= \left( \begin{array}{cc}
     m_2 & \sqrt{2} m_W \sin \beta \\
     \sqrt{2} m_W \cos \beta & -\lambda x  \end{array} \right)
\eea
It is diagonalized by the orthogonal matrices $U,V$ (we do not include
beyond CKM CP violation)
\be
U M_{\chi^\pm} V^T = \rm diag (m_{\chi_1},m_{\chi_2}) 
\ee

The stop mixing matrix is given as
\bea
\left( \begin{array}{c} \tilde t_1 \\ \tilde t_2 \end{array} \right)= 
\left( \begin{array}{cc}
     \cos \theta_{\tilde t} & \sin \theta_{\tilde t} \\
     -\sin \theta_{\tilde t} & \cos \theta_{\tilde t} \end{array} \right)
\left( \begin{array}{c} \tilde t_L \\ \tilde t_R \end{array} \right)
\eea
Here, $\tilde t_{1,2}$ are the mass and $\tilde t_{L,R}$ the gauge
eigenstates.

The loop functions are defined as
\bea
D_2(x,y) &=& \frac{ x \ln x}{(1-x)(x-y)} + (x\leftrightarrow y) \, ,
~~~~~~~ D_2(1,1)=-\frac{1}{2}\\
D_3(x) &=& \frac{x \ln x }{1-x} \, ,  ~~~~~~~ D_3(1)=-1
\eea

\renewcommand{\theequation}{\Alph{section}-\arabic{equation}}
\setcounter{equation}{0}
\section{Decay rates \label{app:decayrates}}

The rate of the light NMSSM pseudoscalar into down-type fermions is
given as
\be
\Gamma (A_1^0 \to \bar f f) =\frac{1}{4 \pi} \frac{G_F}{\sqrt{2}}
\left( \frac{v}{x} \delta_- \right)^2 m_{A_1^0} m_f^2 
\sqrt{1- 4 \frac{m_f^2}{m_{A_1^0}^2}} \cdot r
\label{eq:ffbarrate}
\ee
where $r=1$ for leptons and $r=N_C$ for quarks.
The decay rate into up-quarks is $1/\tan^4 \beta$ suppressed. 
The rate into two photons reads as 
\be
\Gamma (A_1^0 \to \gamma \gamma) 
=\frac{\alpha^2}{8 \pi^3 } \frac{G_F}{\sqrt{2}} m_{A_1^0}^3 | \sum_i I_i |^2
\label{eq:gamgamrate}
\ee
where for $i=e,\mu,\tau$ and $d,s,b$ loops 
$I_i= r Q_i^2 \kappa_i F(\kappa_i) \delta_- v/x$,
$\kappa_i=m_i^2/m_{A_1^0}^2$ and $Q_i$ is the charge of the fermion.
The function $F(\kappa)$ can be seen in \cite{Gunion:1988mf}. 
It assumes the limits
\bea
\label{eq:fla}
\kappa F(\kappa) =\left\{\begin{array}{l}
\hspace{0.3cm} 0\quad   \mathrm{for } \quad  \kappa \ll 1\\
-\frac{1}{2} \quad  \mathrm{for }\quad \kappa \gg 1 \\
-\frac{\pi^2}{8}   \quad \mathrm{for }  \quad \kappa=\frac{1}{4}
\end{array} \right.
\eea
Higgsino loops contribute as $I_{\tilde \chi}= \sqrt{2} U_{i2} V_{i2} 
\lambda m_W/m_\chi \kappa_{\tilde \chi} F(\kappa_{\tilde \chi}) $. 
It follows from Eq.~(\ref{eq:fla})  that near 
$m_{A_1^0} \leq 2 m_\mu$ the $\gamma \gamma$ 
rate is dominated by the muon loop.
Contributions from up-type quarks are suppressed by $1/\tan^4 \beta$. 

The decay rates for inclusive and exclusive $b \to s A_1^0$ FCNCs read as
\bea
\label{eq:bsA}
\Gamma(B \to X_s A_1^0)&=&\frac{G_F^2 |V_{tb} V_{ts}^*|^2}{2^{10} \pi^5}
|C_A|^2 \frac{(m_b^2-m_{A_1^0}^2)^2}{m_b^3} \\
\label{eq:BKA}
\Gamma(B \to K A_1^0)&=&\frac{G_F^2 |V_{tb} V_{ts}^*|^2}{2^{10} \pi^5}
|C_A|^2 \frac{|\vec p_K|}{m_B^2} |f_0(m_{A}^2)|^2 (\frac{m_B^2-m_K^2}{m_b})^2
\eea
where the form factor $f_0$ parametrizes the matrix element
\be
<K(p_K)| \bar s_L b_R| B(p_B)>=\frac{1}{2} (\frac{m_B^2-m_K^2}{m_b}) 
f_0 ((p_B-p_K)^2)
\ee
Here, $\vec p_K$ denotes the three momentum of the Kaon
and $f_0(0)\sim 0.3$ to 0.4 \cite{Ali:1999mm}.

The branching ratio for radiative $\Upsilon$ decays is given as, 
e.g.~\cite{Haber:1987ua} 
\be
\label{eq:upsbr}
\frac{{\cal{B}}( \Upsilon \to A_1^0 \gamma)}{
{\cal{B}}(\Upsilon \to \mu^+ \mu^-)}= 
\frac{G_F m_{\Upsilon}^2}{4 \sqrt{2} \pi \alpha} 
\left( \frac{v}{x} \delta_- \right)^2
\left( 1-\frac{m_{A_1^0}^2}{m_\Upsilon^2} \right) F
\ee
where $F  \sim  1/2$ includes QCD corrections and ${\cal{B}}(\Upsilon(1s) \to \mu^+ \mu^-)=(2.48 \pm 0.06) \%$ \cite{Hagiwara:fs}.

\end{appendix}


\end{document}